\documentclass[aps,prbbib,twocolumn,epsf]{revtex4}
\usepackage{graphicx}

\begin{document}

\draft

\title{Theory of ZT enhancement in nanocomposite materials.}
\author{Paul M. Haney$^{1}$}

\affiliation{$^{1}$Center for Nanoscale Science and Technology,
National Institute of Standards and Technology, Gaithersburg,
Maryland 20899-6202, USA }

\begin{abstract}
The effect of interface scattering on the performance of disordered, nanocomposite thermoelectric materials is studied theoretically using effective medium theory and direct numerics.  The interfacial electronic and phonon scattering properties which lead to an enhancement of the thermoelectric figure of merit $ZT$ are described.  Generally, $ZT$ enhancement requires the interfacial electrical conductance to be within a range of values, and the thermal phonon conductance to be below a critical value.  For the systems considered, these requirements on interface scattering for $ZT$ enhancement are expressed in terms of the bulk properties of the high-$ZT$ material, and the ratio of the constituent bulk $Z$ values.

\end{abstract}

\maketitle
\section{introduction}

There has been considerable recent interest in utilizing nanostructure to enhance thermoelectric performance \cite{review1}.  A good thermoelectric has scattering mechanisms for phonons and electrons with different features:  electron scattering should be strongly energy-dependent, while phonon scattering should simply be strong.  Nanostructured materials may provide a route to meeting both requirements \cite{review2}.  Nanostructure can change a material's basic electronic properties; for example, the inclusion of localized impurity states can enhance peaks in the density of states \cite{heremans}, leading to stronger energy-dependence of scattering.  Alternatively, nanostructure on a length scale greater than the mean free path does not change the constituent materials' basic electronic properties, but scattering at the interface between material phases changes the bulk composite properties.  A mismatch in material density or sound speed generally decreases the phonon conductivity through interface scattering, and some interfaces provide a potential ({\it e.g.} a Schottky barrier) which serves as an effective energy filter, transmitting higher energy electrons, while blocking lower energy electrons \cite{martin}.  The effect of nanostructuring on the thermoelectric figure of merit $ZT$ was systematically studied in Ref. (\onlinecite{poudeu}), where $ZT$ enhancement was observed for a range of nanocomposite mixing.  Previous theoretical works have analyzed in detail the electron \cite{leonard, popescu} and phonon scattering \cite{chen} at specific interfaces.  Interfaces that scatter electrons and phonons as described above may increase $ZT$, and a more quantitative and general description of the required interfacial properties for $ZT$ enhancement in composite materials is provided here.

In this work, I employ a linear response model of transport to study disordered, two-component materials - including the effects of interface scattering - using effective medium theory and direct numerics.  I describe the specific electronic and phonon scattering properties which lead to $ZT$ enhancement of the composite.  The requirements for $ZT$ enhancement are expressed in terms of the bulk properties of the high $ZT$ constituent, and the ratio of the constituent bulk $Z$ values.  Analysis of these requirements demonstrates the challenges with the nanostructuring approach for $ZT$ enhancement, but should facilitate an efficient search for materials that provide higher efficiency.

\section{Model}
The starting point is the linear response description of transport for the electrical current $j$ and thermal current $j_Q$ \cite{footnote}:
\begin{eqnarray}
j &=& -\sigma \nabla V + \sigma S\nabla T ~,\nonumber \\
j_Q &=& -\left(\kappa^e+\kappa^\gamma\right) \nabla T + \sigma S T \nabla
V~,\label{eq:transport}
\end{eqnarray}
\begin{eqnarray}
\nabla \cdot j = 0;~~~\nabla\cdot j_Q = 0,\label{eq:cont}
\end{eqnarray}
where $\sigma$ is the local electrical conductivity, $\kappa^e~\left(\kappa^\gamma\right)$ is the electron (phonon) contribution to the total local thermal conductivity $\kappa$ ($\kappa=\kappa^e+\kappa^\gamma$) (all thermal conductivities evaluated for zero electric field), $S$ is the thermopower, $V$ is the electrostatic potential, and $T$ is the temperature.  I assume that $\sigma$ and $\kappa^e$ obey the Wiedamann-Franz law: $\kappa_e=\sigma L_0T$, where $L_0$ is the Lorenz number. Eq. (\ref{eq:transport}) is valid for length scales greater than a mean free path, which for relevant materials is on the order of 10 nm.

The figure of merit $ZT$ is:
\begin{eqnarray}
ZT &=&
\frac {S^2 \sigma T}{\kappa -  S^2\sigma T} = \frac{N}{1-N + K}~.
\label{eq:zt}
\end{eqnarray}
where $K=\left(\kappa^\gamma/\kappa^e\right)$ and $N=S^2\sigma T/\kappa^e$.  $N$ can be rewritten in terms of the thermopower only, using the Wiedamann-Franz law: $N=S^2/L_0$.  $N$ is constrained by the second law of thermodynamics to be less than 1.  Equivalently, $S$ is always less than $\sqrt{L_0}\equiv S_{\rm max}$.  An ideal thermoelectric has electronic properties such that $N\rightarrow 1$, and low phonon thermal conductivity such that $K\rightarrow 0$.

\begin{figure}[h!]
\begin{center}
\vskip 0.2 cm
\includegraphics[width=3.4in]{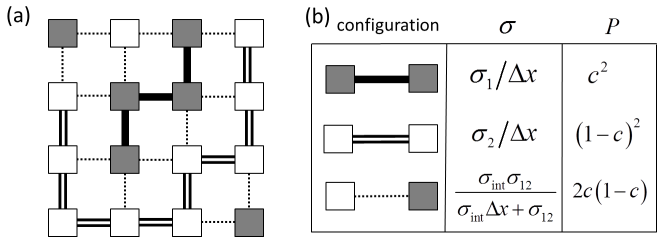}
\vskip 0.2 cm \caption{(a) depicts a typical random site configuration, where the links between sites are set by the adjacent site types. (b) shows the values of conductance for each link type, along with the probability for each link type.  As discussed in the text, $\Delta x$ is the grain size, and $\sigma_{12}$ is the series conductivity of $\sigma_1$ and $\sigma_2$.}\label{fig:resistors}
\end{center}
\end{figure}

To study the thermoelectric properties of nanocomposites, I solve Eqs. (\ref{eq:transport}-\ref{eq:cont}) directly for an ensemble of randomly disordered configurations in 3-d.  Fig. (\ref{fig:resistors}) shows a schematic of the method.  I use a random site approach in which sites are randomly assigned as material 1 with probability $c$, or material 2 with probability $(1-c)$.  The link between two sites represents a resistor (or conductance), whose value is set by the adjacent site types.  Fig. (\ref{fig:resistors}) shows the conductance values for the three possible cases, along with the associated probability for each case.  In the table, $\sigma_1$ ($\sigma_2$) is the bulk conductivity for material 1 (material 2), and $\sigma_{12}=\sigma_1\sigma_2/\left(\sigma_1+\sigma_2\right)$.  $\sigma_{\rm int}$ is the interface conductance, and $\Delta x$ is the grain size of the material.  In the absence of interface scattering, $\Delta x$ factors out of the problem and is not important.  In the presence of interface scattering, $\Delta x$ is a key parameter: a small $\Delta x$ implies a higher interface density, and a more significant effect of the interface scattering.  It's important to note that this theory applies only to materials for which $\Delta x$ is greater than the mean free path.  Finally, I note that this scheme is not unique; Appendix B discusses more complicated schemes, and shows comparisons between different schemes.  The advantage of the simple approach described here is that it captures the physics of the systems studied well, and is amenable to analytic treatment with effective medium theory.

Numerically, the system is discretized into $20^3$ sites (the results do not change appreciably when going to $30^3$ sites), and the ensemble size is such that the statistical error of the effective transport parameters is converged (this typically requires about 30 configurations).  The error bars on the plots of numerical results indicate the statistical uncertainty (one standard deviation).

The transport properties of a multi-component, disordered system can be approximated with effective medium theory (EMT). As shown in Ref. (\onlinecite{cohen}), the effective medium electrical conductivity $\sigma$, total thermal conductivity $\kappa$, and thermopower $S$ satisfy:

\begin{eqnarray}
\sum_i P_i \left(\frac{\sigma_i-\sigma}{\sigma_i+2\sigma}\right)
=\sum_i P_i \left(\frac{\kappa_i-\kappa}{\kappa_i+2\kappa}\right)&&
=0~,\label{eq:kappatotemt}
\end{eqnarray}

\begin{eqnarray}
& S=&3\kappa\sigma
\left(\sum_i P_i \frac{\sigma_i S_i}{\left(\kappa_i+2\kappa
\right)\left(\sigma_i+2\sigma\right)}\right)\times
\nonumber
\\ &&\left(\sum_i P_i \left[\frac{\sigma_i\kappa+\sigma\kappa_i+2\sigma\kappa-\sigma_i\kappa_i}{\left(\kappa_i+2\kappa
\right)\left(\sigma_i+2\sigma\right)}\right]\right)^{-1},\label{eq:pemt}
\end{eqnarray}
where $i$ labels the link type, and $P_i$ is the probability of a link with transport parameter values $\sigma_i,~\kappa_i,~S_i$.

Each material type (bulk 1, bulk 2, and interface) is described by three parameters: $\left(\sigma,\kappa^\gamma,S\right)$, so that 9 material parameters (plus the concentration $c$) describe a specific two-component system.  This parameter space is too large to describe in its entirety.  To make progress, I generally present results for fixed bulk properties, fixed interfacial thermopower, and vary the interfacial electrical and phonon thermal conductivities.

Appendix A discusses the scaling of Eqs. (\ref{eq:transport}) to dimensionless form.  The transport coefficients $\left(\sigma,\kappa^\gamma,S\right)$ end up being scaled by those of material 1 (so that $\sigma_2 \rightarrow \left(\sigma_2/\sigma_1\right)$; the interface values also have a value of $\Delta x$ present in their dimensionless form: $\sigma_{\rm int} \rightarrow \left(\sigma_{\rm int}/\sigma_1\right)\Delta x$).  For ease of presentation, I omit this explicit scaling in most plots; the axis label $\bar{\sigma}_{\rm int}$ refers to $\left(\sigma_{\rm int}/\sigma_1\right)\Delta x$ , and the label $K_{\rm int}$ refers to $\left(\kappa_{\rm int}^{\gamma}/\kappa_1^e\right)\Delta x$.

\section{Results}

\subsection{single interface}

To illustrate the qualitative effect of interface scattering on thermoelectric performance - and the conditions under which $ZT$ is enhanced - it suffices to consider the simplest possible system: 1-d transport in a bilayer.  This maps onto a 3-resistor-in-series problem.

\begin{figure}[h!]
\begin{center}
\vskip 0.2 cm
\includegraphics[width=3.2in]{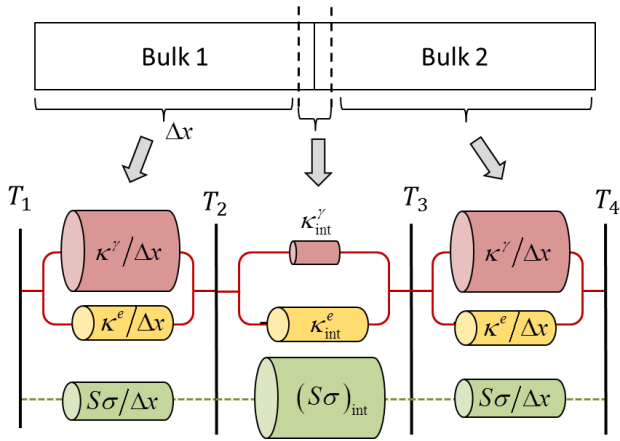}
\vskip 0.2 cm \caption{Depiction of the scattering in a simple bilayer.  The phonon thermal conductance (red cylinder) is detrimental to thermoelectric performance, while the thermoelectric conductance (green cylinder) is beneficial.  Interface scattering can improve overall thermoelectric performance by improving either or both of these transport processes, as shown in the figure.  Here $\Delta x$ refers to the layer thickness.}\label{fig:interface}
\end{center}
\end{figure}

Fig. (\ref{fig:interface}) illustrates the role of interface scattering in increasing $ZT$.  The solid red lines denote paths for heat current (top red line for phonons, bottom red line for electrons), the green dashed line for thermoelectric charge current.  The cylinder size represents the magnitude of the conductance for a specific transport path.  Interface scattering can increase $ZT$ in two ways: 1. by reducing the phonon thermal conductivity, which leads to $K\rightarrow 0$, or 2. by increasing the thermopower, which leads to $N\rightarrow 1$.  The interface conductances in Fig. (2) improve $ZT$ in both ways.  In the rest of the paper, I focus on the scenario in which $ZT$ is enhanced via increased phonon scattering.  One reason for this is that enhancement via increase in thermopower is less well described by this numerical model.  See Appendix B for further discussion on this point.

\begin{figure}[h!]
\begin{center}
\vskip 0.2 cm
\includegraphics[width=3.4in]{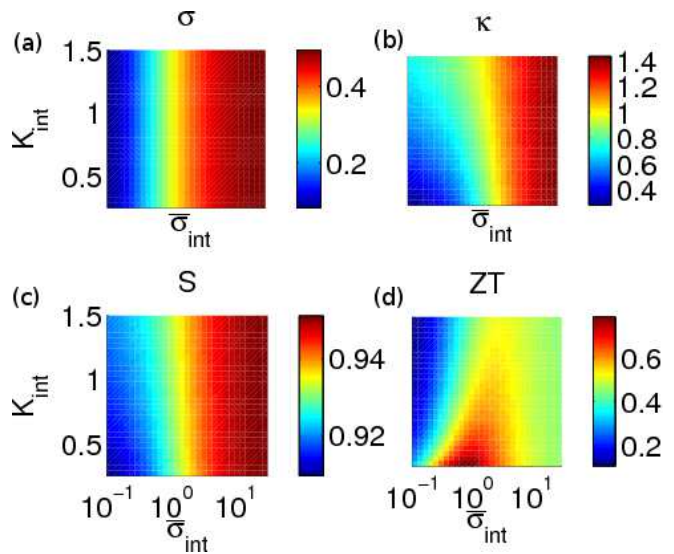}
\vskip 0.2 cm \caption{Transport parameters of bilayer as $\bar{\sigma}_{\rm int}$ and $K_{\rm int}$ are varied. (Recall the axes labels omit scaling factors.  Their inclusion is via:  $\bar{\sigma}_{\rm int} = \left(\sigma_{\rm int}/\sigma_1\right)\Delta x$ and $K_{\rm int}= \left(\kappa_{\rm int}^{\gamma}/\kappa_1^e\right)\Delta x$.)  The fixed system parameters are: $\sigma_2=\sigma_1,~\kappa_2^\gamma=\kappa_1^\gamma,~S_1=S_{\rm max},~S_2=0.9~S_{\rm max},~S_{\rm int} = 0.9~ S_{\rm max}$ (so that $Z_1T=0.5,~Z_2T=0.375$). }\label{fig:3layer}
\end{center}
\end{figure}

Fig. (\ref{fig:3layer}) shows the transport properties of the layer for fixed bulk properties, and varying the interface electrical conductance and phonon thermal conductance (the interface thermopower is fixed).  The results are intuitively clear: when the interface electrical conductance is small, it determines the overall layer conductance; conversely when the interface electrical conductance is large, the interface is transparent and the overall conductance is set by the bulk.  A similar scenario holds for the thermal conductance (though now the total thermal conductance depends on both electrical and phonon components).  I've assumed the thermopower is high for all constituents, so that its value is relatively unaffected by the interface.  This leads to a $ZT$ value which is enhanced relative to the bulk for a certain range of interfacial transport parameters.

\begin{figure}[h!]
\begin{center}
\vskip 0.2 cm
\includegraphics[width=3.2in]{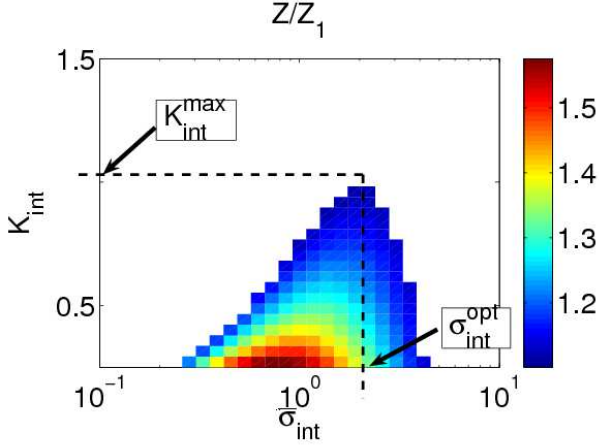}
\vskip 0.2 cm \caption{Replot of Fig. (\ref{fig:3layer}d): $Z$ of the trilayer normalized by $Z$ of the high-$ZT$ bulk constituent. Only values for which $ZT$ of the trilayer is 5\% greater than the bulk are shown.  I use the parameters $K_{\rm int}^{\rm max},~\sigma_{\rm int}^{\rm opt}$ to characterize the phase space of interface properties that lead to $ZT$ enhancement. }\label{fig:ztsimple}
\end{center}
\end{figure}

The region of $ZT$ enhancement is shown again in Fig. (\ref{fig:ztsimple}), where only values for which $ZT$ is more than 5\% greater than the bulk value are shown.  For disordered materials in 2 and 3 dimensions, the phase space of $ZT$ enhancement is very similar to this 3-resistor case, so it's worth investigating this simple example fully.

In the limit of low interface conductance (the lower left-hand portion of Fig. (\ref{fig:ztsimple})), the interface properties dominate.  $ZT$ of the layer is then approximately that of the interface, so the contours of Fig. (\ref{fig:ztsimple}) in this region are simply those of Eq. (\ref{eq:zt}), with $N\rightarrow N_{\rm int},~K\rightarrow K_{\rm int}$.  The small $\sigma_{\rm int}$ in this region implies a small electron thermal conductance $\kappa^e_{\rm int}$, via the Wiedamann-Franz law.   A large $ZT$ then requires a very small phonon thermal conductance $\kappa_{\rm int}^\gamma$, making $ZT$ enhancement in this region difficult to achieve.  (Recall that for high thermopower, $ZT$ is set by the ratio of $\kappa^e$ to $\kappa^\gamma$, see Eq. (\ref{eq:zt}).) In the opposite limit of high interface conductance ($\bar{\sigma}_{\rm int} \gg 1$), the interface is transparent electrically and thermally  (thermal transparency follows from Wiedamann-Franz law: $\sigma_{\rm int} \rightarrow \infty \Rightarrow \kappa^e_{\rm int} \rightarrow \infty$).  Here purely bulk properties are recovered, and $ZT$ is not increased.

The crossover between these limits occurs around $\bar{\sigma}_{\rm int}=1$ (or $\sigma_{\rm int} = \sigma_1/\Delta x$), when both interface and bulk properties are important.  This is the region most accessible for $ZT$ enhancement.  Not surprisingly, $ZT$ is always increased as the phonon thermal conductance of the interface is decreased.  I label the maximum value of $K_{\rm int}$ for which there is a $ZT$ enhancement of 5\% over the bulk value as $K_{\rm int}^{\rm max}$.  (Recall this parameter in full scaled form is $K_{\rm int}^{\max}=\left(\kappa^\gamma_{\rm int}/\kappa^e_1\right)\Delta x$.)  This is a key parameter because finding materials with interface scattering that lies {\it below} this value is a primary challenge for using nanocomposites for $ZT$ enhancement \cite{review}.  I label the associated electrical conductance $\sigma_{\rm int}^{\rm opt}$ (see Fig. (\ref{fig:ztsimple})).  The next section is largely devoted to describing how the values of $K_{\rm int}^{\rm max}$ and $\sigma_{\rm int}^{\rm opt}$ depend on the properties of the bulk material constituents.

\begin{figure}[h!]
\begin{center}
\vskip 0.2 cm
\includegraphics[width=3.4in]{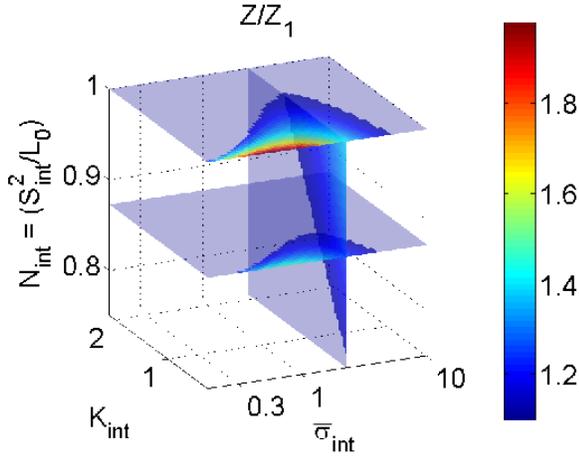}
\vskip 0.2 cm \caption{Region of $ZT$ enhancement for the full parameter space of the interface.  Interfaces with high $N$ (high thermopower) are advantageous for $ZT$ enhancement.  The same bulk parameters are used as in Fig. (\ref{fig:3layer}).}\label{fig:intspace}
\end{center}
\end{figure}

So far I have fixed the interface thermopower $S_{\rm int}$.  To illustrate how the space of $ZT$ enhancement depends on $S_{\rm int}$, I make some slices through the full parameter space of the interface, shown in Fig. (\ref{fig:intspace}).  Not surprisingly, as the thermopower of the interface decreases, the space of $ZT$ enhancement in $\left(\bar{\sigma}_{\rm int},~K_{\rm int}\right)$ gets smaller ({\it i.e}, it's harder to achieve enhancement when the interfacial thermopower is weak).  In the rest of the paper, I fix $S_{\rm int}=S_{\rm max}$ (or $N_{\rm int}=1$).  It should be kept in mind that an interface with smaller $S$ will have more stringent requirements on phonon thermal conductance ({\it i.e} a smaller $K_{\rm int}^{\rm max}$) for $ZT$ enhancement.

\subsection{3-d disordered material}

\begin{figure}[h!]
\begin{center}
\vskip 0.2 cm
\includegraphics[width=3.4in]{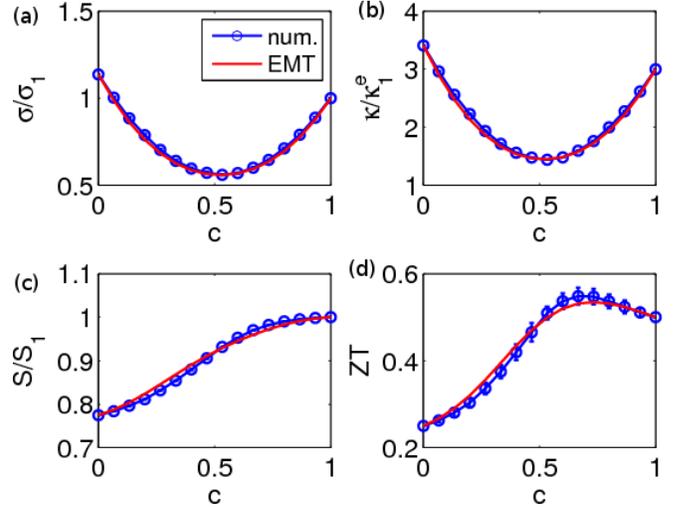}
\vskip 0.2 cm \caption{The transport parameters of a two-component disordered medium as a function of relative concentration.  System parameters are: $\sigma_2 = 1.1~ \sigma_1,~\kappa_1^\gamma= 2~ \kappa_1^e,~\kappa_2^\gamma = 2.3~ \kappa_1^e, S_1=S_{\rm max}, S_2 = 0.77 ~S_{\rm max},~ \sigma_{\rm int} = 0.24 ~\sigma_1/\Delta x, ~\kappa_{\rm int}^\gamma = 0.48 ~\kappa_1^e/ \Delta x, ~S_{\rm int} = 0.97 ~S_{\rm max}$.  (a) and (b) show a decrease in the conductance due to interface scattering.  (d) shows an enhancement in $ZT$.}\label{fig:3d_4panel}
\end{center}
\end{figure}

Moving to disordered materials in 3-d introduces a new system parameter - the concentration of one material with respect to the other.  Fig. (\ref{fig:3d_4panel}) shows the bulk transport parameters as a function of concentration calculated numerically and with effective medium theory.  The interface scattering leads to a decrease in electric and thermal conductivity relative to the bulk values of the constituents.  $ZT$ is enhanced relative to the bulk value for a range of concentrations, shown in Fig. (\ref{fig:3d_4panel}d).  Note there is excellent agreement between effective medium theory and the numeric results; most of the results presented in the rest of the paper are derived from effective medium theory, except where explicitly noted.

The rest of this section describes how the interface properties needed for $ZT$-enhancement depend on the constituent bulk materials.  I show that the phase space for $ZT$ enhancement essentially depends only on a small number of parameters of the constituent materials.  This is a useful simplification.  It enables a precise answer to the question: ``given a high-$ZT$ material with thermopower $S_1$ and phonon thermal conductivity $\kappa_1^\gamma$, and another material with $Z$ value $Z_2$, what are the interface scattering properties that are required to observe $ZT$ enhancement of the composite material?".  I describe the required interface properties using the parameters $(K_{\rm int}^{\rm max}, \sigma_{\rm int}^{\rm opt})$ introduced in the previous section and in Fig. (\ref{fig:ztsimple}).

For each set of bulk and interface properties, I vary the concentration and determine the maximum possible $ZT$ - this maximum value is what is reported in the following results.  In all of these results, I assume that the bulk thermopower of the high-$ZT$ constituent is large ($S_1=S_{\rm max}$), so that $ZT$ enhancement is a consequence of reducing phonon thermal conductivity.

\begin{figure}[h!]
\begin{center}
\vskip 0.2 cm
\includegraphics[width=3.4in]{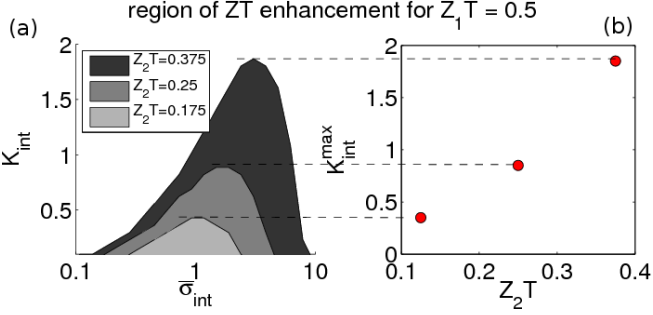}
\vskip 0.2 cm \caption{(a) shows regions of ZT enhancement with respect to interface properties for $Z_1T=0.5$ ($N_1=1,~K_1=2$), $Z_2T$ is decreased by increasing $K_2$, with values (2.67, 4.0, 8.0) ($N_2=1$ for all cases).  (b) shows how this phase plot is translated to a plot of $K_{\rm int}^{\rm max}$ versus $Z_2 T$. }\label{fig:enhancement}
\end{center}
\end{figure}

Fig. (\ref{fig:enhancement}) is an illustration of how $K_{\rm int}^{\rm max}$ characterizes the phase space of $ZT$ enhancement as bulk materials change.  Fig. (\ref{fig:enhancement}a) shows how the region of enhancement changes as the $ZT$ value of one bulk material component gets smaller.  As one component's $ZT$ value decreases, it's more difficult to achieve $ZT$ enhancement of the composite via interface scattering.  Fig. (\ref{fig:enhancement}b) shows how this behavior is translated into the parameter $K_{\rm int}^{\rm max}$.  In this example, $ZT$ of material 2 is degraded due to a higher phonon thermal conductivity of material 2.

\begin{figure}[h!]
\begin{center}
\vskip 0.2 cm
\includegraphics[width=3.5in]{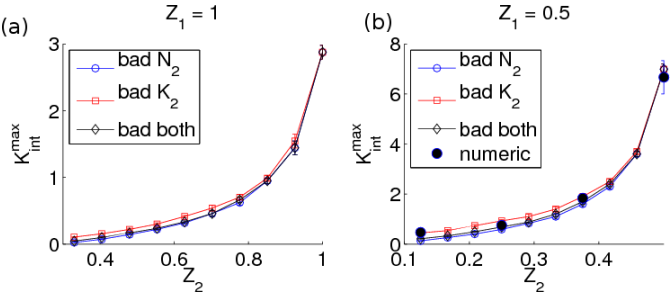}
\vskip 0.2 cm \caption{In (a), $ZT_1=1$ ($N_1=1,~K_1=1$), and $ZT_2$ is reduced in three ways: by decreasing $N_2$, increasing $K_2$, or a combination of both.  (b) shows the same plot, with $ZT_1=0.5$ ($N_1=1,~K_1=2$), and also shows results obtained numerically.}\label{fig:kappa_2cases}
\end{center}
\end{figure}

Fig. (\ref{fig:kappa_2cases}) shows that for a fixed high-$ZT$ constituent, $K_{\rm int}^{\rm max}$ essentially only depends on $Z_2$.  In Fig. (\ref{fig:kappa_2cases}a) I show plots of $K_{\rm int}^{\rm max}$ as $Z_2T$ is degraded in three different ways: with a ``bad" $K_2$ (or high phonon thermal conductivity), a ``bad" $S_2$ (or low thermopower), and a combination of both.   Fig. (\ref{fig:kappa_2cases}b) shows the same thing for a different high-$ZT$ material.  Fig. (\ref{fig:kappa_2cases}b) also shows numeric results (with a ``bad $K_2$" scenario), which confirm that the effective medium theory and numerical results are very similar.  What's important is that $K_{\rm int}^{\rm max}$ is quite insensitive to {\it how} the low-$ZT$ material is deficient.  The bulk $ZT$ values of the constituent alone determines the required interfacial phonon scattering for $ZT$ enhancement\cite{footnote1}.

\begin{figure}[h!]
\begin{center}
\vskip 0.2 cm
\includegraphics[width=2.5in]{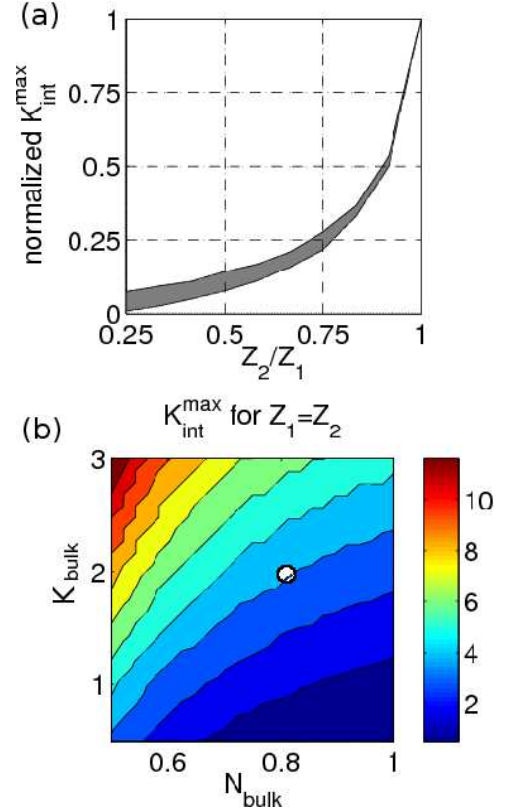}
\vskip 0.2 cm \caption{(a) shows the range of $K_{\rm int}^{\rm max}$ values as a function of $Z_2/Z_1$ (from Fig. 8), when normalized by their maximum value.  (b) shows this overall normalization constant as a function of the thermoelectric parameters $N$ and $K$ of the high-$ZT$ bulk material.  The white dot refers to an example described in the text.}\label{fig:kappamax}
\end{center}
\end{figure}

Fig. (\ref{fig:kappamax}a) shows the result of plotting all the curves of Fig. (\ref{fig:kappa_2cases}) together, normalized by their maximum value.  Again a remarkable and useful simplification takes place, where the curves collapse on an approximately ``universal" curve.  The vertical spread of this normalized curve shows the spread in values for the different curves of Fig. (\ref{fig:kappa_2cases}).

The right hand-side of the normalized curve of Fig. (\ref{fig:kappamax}a), where $Z_2=Z_1$, corresponds to a system with identical bulk phases, with interface scattering between the identical grains.  The value of $K_{\rm int}^{\rm max}$ for such a system sets the overall normalization for plots like Fig. (\ref{fig:kappa_2cases}).  In Fig. (\ref{fig:kappa_2cases}b), I plot the value of this normalization as a function of $K_{\rm bulk}$ and $N_{\rm bulk}$. The two parts of Fig. (\ref{fig:kappamax}) enable an estimate for the required interface phonon thermal conductance for $ZT$ enhancement.

As an example, let the high-$ZT$ constituent have $N_1=0.8$, $K_1=2$ (this implies $Z_1T=0.36$); this is shown as a white dot in Fig. (\ref{fig:kappamax}).  Fig. (\ref{fig:kappamax}b) shows the normalization for the $K_{\rm int}^{\rm max}$ curve is 4.  Now let the low-$ZT$ material have $Z_2T=0.27$, so that $Z_2/Z_1=0.75$.  Using Fig. (\ref{fig:kappamax}a), I conclude $K_{\rm int}^{\rm max}$ for this material combination is $0.25\times 4 = 1$ (in dimensionful terms, $K_{\rm int}^{\rm max}=\left(\kappa_{\rm int}^\gamma/\kappa_1^e \right)\Delta x = 1$).  This means that $ZT$ enhancement requires thermal transport parameters and grain size such that $\left(\kappa_{\rm int}^\gamma/\kappa_1^e \right)\Delta x < 1$.

\begin{figure}[h!]
\begin{center}
\vskip 0.2 cm
\includegraphics[width=3.2in]{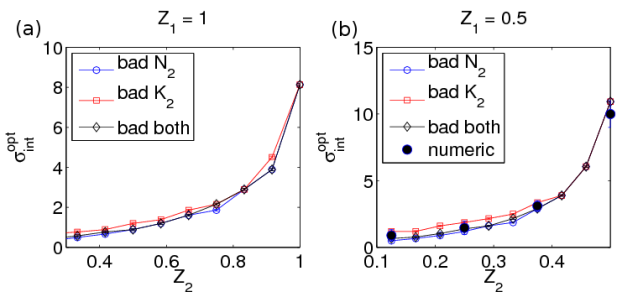}
\vskip 0.2 cm \caption{In (a), $ZT_1=1$ ($N_1=1,~K_1=1$), and $ZT_2$ is reduced in three ways: by decreasing $N_2$, increasing $K_2$, or a combination of both.  (b) shows the same plot, with $ZT_1=0.5$ ($N_1=1,~K_1=2$), and also shows results obtained numerically.}\label{fig:sigma_2cases}
\end{center}
\end{figure}

\begin{figure}[h!]
\begin{center}
\vskip 0.2 cm
\includegraphics[width=2.5in]{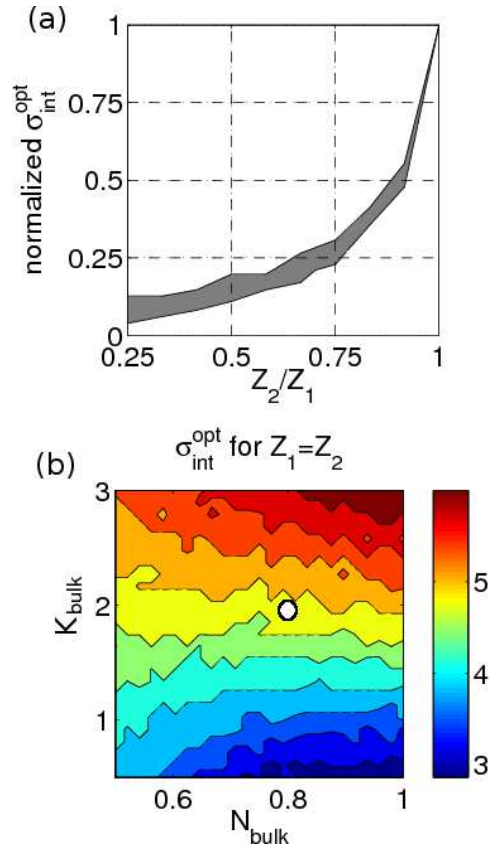}
\vskip 0.2 cm \caption{(a) shows the range of $\sigma_{\rm int}^{\rm opt}$ values as a function of $Z_2/Z_1$ (from Fig. 10), when normalized by their maximum value.  (b) shows this overall normalization constant as a function of the thermoelectric parameters $N$ and $K$ of the high-$ZT$ bulk material.}\label{fig:sigmaOpt}
\end{center}
\end{figure}

I next go through a similar description of how $\sigma_{\rm int}^{\rm opt}$ depends on bulk material parameters.  Fig. (\ref{fig:sigma_2cases}) shows $\sigma_{\rm int}^{\rm opt}$ as a function of the $Z_2/Z_1$, for two different high-$ZT$ constituents.  I again find that, for fixed high-$ZT$ material, $\sigma_{\rm int}^{\rm opt}$ essentially only depends on $Z_2$.  Fig. (\ref{fig:sigmaOpt}a) shows the result of plotting all the curves of Fig. (\ref{fig:sigma_2cases}) by their maximum value. Again, I find that $\sigma_{\rm int}^{\rm opt}$ as a function of $Z_2/Z_1$ is an approximately ``universal" curve.  Fig. (\ref{fig:sigmaOpt}b) shows the normalization value of this ``universal" curve as a function of $N$ and $K$ of the high-$ZT$ constituent.  These plots again enable an estimate of $\sigma_{\rm int}^{\rm opt}$ in terms of just a few bulk material parameters.

Returning to the example before (where we assumed a high $ZT$ material with parameters $N_1=0.8,~K_1=2$, and a low $ZT$ material with $Z_2/Z_1=0.75$),
Fig. (\ref{fig:sigmaOpt}b) shows the normalization constant of about 5, which is used with Fig. (\ref{fig:sigmaOpt}a) to infer $\sigma_{\rm int}^{\rm opt} = 5\times0.25=1.25$.  In other words, attaining $ZT$ enhancement through interfacial scattering is most easily accessible with a combination of electrical conductivity values and grain size such that $\left(\sigma_{\rm int}/\sigma_1\right)\Delta x = 1.25$.

$\sigma_{\rm int}^{\rm opt}$ is an important constraint on the interface; even if an interface blocks phonons effectively, if it also blocks electrons too much ({\it i.e} has too low $\bar{\sigma}_{\rm int}$), or is transparent to electrons (too high $\bar{\sigma}_{\rm int}$), then it does not lead to overall $ZT$ enhancement.  The reason for this is the same as in the simple 3-resistor-in-series case, described earlier.  Note that the value of the overall normalization for $\sigma_{\rm int}^{\rm opt}$ is fairly constant over the range of bulk material parameters.  Generally, $ZT$ enhancement requires an interface conductance on the order of the bulk conductivity divided by the grain size.

Figs. (\ref{fig:kappamax}) and (\ref{fig:sigmaOpt}) represent the main results of the paper.  They provide a blueprint to choosing material properties such that a two-component composite results in $ZT$ enhancement.  An important aspect of Fig. (\ref{fig:kappamax}a) is the rapid decrease of $K_{\rm int}^{\rm max}$ as one of the material's $ZT$ value decreases.  This means interfacial phonon scattering can most easily enhance $ZT$ when the two materials have similar $ZT$ values.  This poses a key materials science challenge in pursuing this technique for $ZT$ enhancement: often materials with similar (high) $ZT$ values have similar density ({\it i.e.} both composed of heavy atoms); however, interfacial phonon scattering is usually strongest between materials with very dissimilar density and speed of sound \cite{review}.

For a rough estimate of required material values, the above analysis shows $ZT$ enhancement via interface phonon scattering requires material parameters which satisfy an inequality on the order of $\kappa_{\rm int}^\gamma < \kappa_{\rm bulk}^e / \Delta x$.  A typical thermoelectric has $\kappa_{\rm bulk}^e = 1~ {\rm W/\left(m\cdot K\right)}$.  Assuming a grain size of $10~{\rm nm}$, the interface phonon conductance must be less than $10^8~{\rm W/\left(m^2\cdot K\right)}$ for $ZT$ enhancement.  This value is certainly attainable for some material combinations \cite{review}, though obtaining this value of $\kappa_{\rm int}^\gamma$ for two materials with high $ZT$ values (and low $\kappa_{\rm bulk}^\gamma$ values) is likely to be a challenge.

\subsection{Dimension and concentration dependence}

\begin{figure}[h!]
\begin{center}
\vskip 0.2 cm
\includegraphics[width=3.4in]{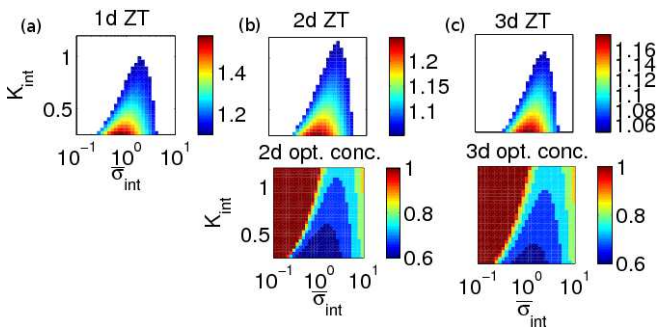}
\vskip 0.2 cm \caption{(a-c) show the region of $ZT$ enhancement in 1, 2, and 3 dimensions (1d refers to the bilayer case).  Below the 2-d and 3-d cases, the concentration with the maximum $ZT$ is shown (concentration refers to percentage of material 1).  Fixed system parameters in all cases are: $N_1=N_2=1$, $K_1=K_2=2$, $S_1=S_{\rm max},~S_2=0.9S_{\rm max},~S_{\rm int}=0.9S_{\rm max}$. }\label{fig:dimensions}
\end{center}
\end{figure}

Here I briefly compare the results obtained for the space of $ZT$ enhancement in 1-d, 2-d, and 3-d.  The comparison is shown in Fig. (\ref{fig:dimensions}).  The interface parameter space for enhancement is very similar in all cases, but that the enhancement is reduced in higher dimensions.  This is because some portion of interface scattering in higher dimensions occurs in directions orthogonal to the transport direction.  This scattering is not effective in reducing the phonon thermal conductivity along the overall direction of the temperature gradient, and therefore does not aid in increasing $ZT$.  Also shown in Fig. (\ref{fig:dimensions}) is the concentration in 2-d and 3-d for which the maximum $ZT$ occurs.  This value depends on the specific material parameters chosen.  For example. if the two bulk materials are equivalent, the optimum enhancement is always at $c=0.5$.  As the two bulk materials properties deviate, the optimum concentration moves away from $0.5$ - it's more advantageous to have a higher concentration of the high-$ZT$ material.  At the edge of the phase space of enhancement, the optimum concentration is such that the composite is mostly high-$ZT$ bulk.

\section{Conclusion}

In this work I described the conditions under which the formation of a nanocomposite material results in enhancement of $ZT$ over the constituent bulk values.  $ZT$ enhancement is the result of electronic and phonon scattering at the interface between different materials, and occurs over a range of $\bar{\sigma}_{\rm int}$, and for sufficiently low $K_{\rm int}$.  Using effective medium theory and numerical simulation, I give a prescription for the required value of interface conductances for $ZT$ enhancement, as a function of the bulk $N$ and $K$ of the high $ZT$ material, and the ratio of the bulk $Z$ values.  The results presented in the 3-d disordered case are for $S_{\rm int}=S_{\rm max}$, and therefore represent the most optimistic requirements on $K_{\rm int}$ and $\bar{\sigma}_{\rm int}$.

I emphasize that this theory applies for composites with phase separation greater than the mean free path of electrons and phonons.  It's therefore most applicable to nanostructuring techniques such as ball milling and hot pressing.  These techniques have shown the potential for $ZT$ enhancement \cite{martin2,poudel}.  Although not emphasized in this work, scattering at interfaces can also improve efficiency via improved energy filtering, resulting in enhanced power factor.  The material constraints to achieve $ZT$ enhancement are obviously challenging, but the precise specification of these constraints should aid in the search for the best material choices for more efficient thermoelectrics.


\section{Appendix}

\subsection{Dimensionless variables}

To write Eqs. (\ref{eq:transport}) in dimensionless form, I introduce the following variables.
\begin{eqnarray}
\overline{x} = \frac{x}{L};
&\overline{\nabla} = L\nabla;&\\
\overline{T} = \frac{T}{T_0}; &
~~\overline{V} = V \left(\frac{S_1 \sigma_1}{\kappa_1^e}\right);&\\
\overline{j} = j \left(\frac{L}{S_1 \sigma_1 T_0}\right); &
~~\overline{j_Q} = j_Q \left(\frac{L}{\kappa_1^e T_0}\right),&
\end{eqnarray}
where $L$ is the length of the sample in the transport direction, $T_0$ is a fixed reference temperature.
This leads to the dimensionless equations:
\begin{eqnarray}
\overline{j} &=& -\frac{1}{N_1}\left(\frac{\sigma_i}{\sigma_1}\right)
\overline{\nabla}~ \overline{V} + \left(\frac{S_i \sigma_i}{S_1 \sigma_1}\right) \overline{\nabla} ~\overline{T}\nonumber \\
\overline{j_Q} &=& -\left(\frac{\kappa_i}{\kappa_1}\right) \overline{\nabla}~\overline{ T} +\left( \frac{S_i \sigma_i}{S_1 \sigma_1}\right) \overline{T} ~ \overline{\nabla}~\overline{V},
\end{eqnarray}\label{eq:dd_dim3}
where $N_1=\left(\frac{S_1^2 \sigma_1 T_0}{\kappa_1^e}\right)$.  The prefactor $1/N_1$ of the dimensionless conductivity results in an ``effective" conductivity $\frac{1}{N_1}\left(\frac{\sigma_i}{\sigma_1}\right)$ that is used when solving Eqs. (\ref{eq:dd_dim3}).  Extracting an effective conductivity from evaluating the charge current response to an electric potential requires accounting for $N_1$: $\sigma = N_1 \left(\frac{j}{\Delta V}\right)$, where $\Delta V$ is the applied potential difference.

\subsection{Discretization scheme}

\begin{figure}[h!]
\begin{center}
\vskip 0.2 cm
\includegraphics[width=3.4in]{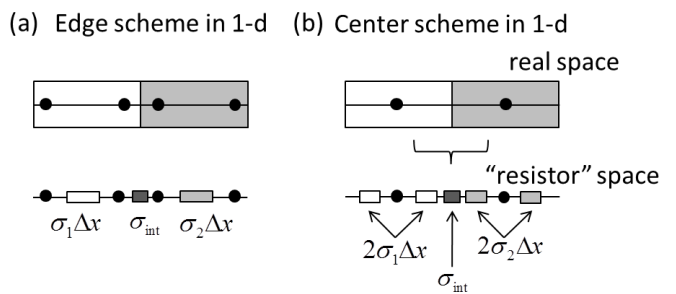}
\vskip 0.2 cm \caption{Two different discretization schemes represented in 1-d.  In the ``center scheme" (a), the interface conductance is partially combined with bulk conductances, and the potential is evaluated at the center of each plaquette.  In the ``edge scheme" (b), the conductances are separate and the potential is evaluated at both edges of the plaquettes.}\label{fig:d1d}
\end{center}
\end{figure}

\begin{figure}[h!]
\begin{center}
\vskip 0.2 cm
\includegraphics[width=3.4in]{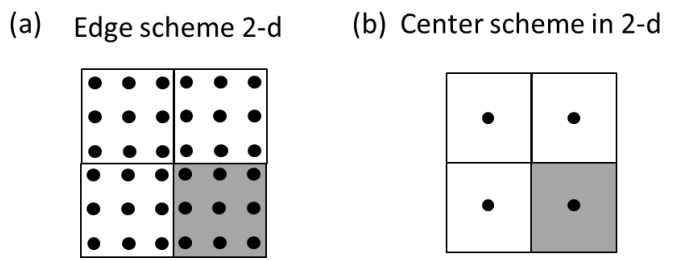}
\vskip 0.2 cm \caption{Implementation of (a) center and (b) edge schemes in 2-d.}\label{fig:d2d}
\end{center}
\end{figure}

The inclusion of interface scattering complicates the scheme used to discretize Eqs. (\ref{eq:transport}-\ref{eq:cont}), which we discuss more fully here.  The relevant question is: given a continuous distribution of material, what discrete set of points should we choose to represent the potential and temperature fields?  The answer depends on the spatial variation of the fields; to accurately represent the continuous fields requires a more dense mesh near areas of rapid variation in potential and temperature.  For example, small interface electrical conductance (compared to the bulk conductivity divided by grain length) implies a sharp potential drop across an interface.  This suggests a discretization scheme as shown in Fig. (\ref{fig:d1d}a).  The conductance on the link separating two plaquettes is set to $\sigma_{\rm int}$ for plaquettes with different identities, and set to $\infty$ otherwise.  I call this discretization scheme the ``edge scheme".  In two dimensions the sampling may be chosen as shown in Fig. (\ref{fig:d2d}a).

\begin{figure}[h!]
\begin{center}
\vskip 0.2 cm
\includegraphics[width=3.4in]{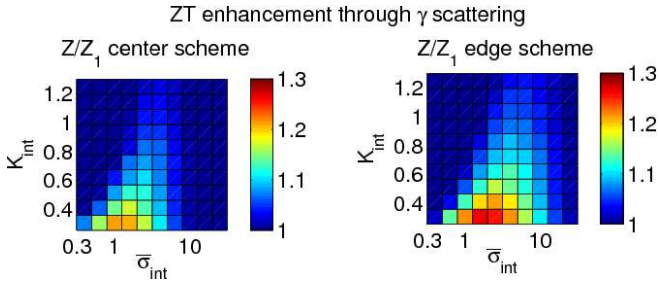}
\vskip 0.2 cm \caption{$ZT$ of the composite versus interface $\bar{\sigma}_{\rm int}$ and $K_{\rm int}$.  The system parameters are: $40{\rm x}40$ plaquettes in 2-d, $N_1=N_2=1$, $K_1=K_2=2$ (so that $Z_1 T=Z_2 T = 0.5$), $S_{\rm int} = 0.9~S_{\rm max}$.  Interface scattering of phonons reduce $K$ of the composite, resulting in an enhancement of $ZT$.}\label{fig:k}
\end{center}
\end{figure}

In the body of the paper I use a simpler scheme, depicted in Fig. (\ref{fig:d1d}b).  Here the fields are evaluated at the center of the plaquette, and the interface conductance is put in series with the adjacent bulk conductance {\it a priori}.  I call this the ``center scheme".  This results in a less dense sampling, and is therefore not as accurate as the edge scheme.  However, as mentioned in the body of the paper, this scheme is easily adopted to effective medium theory, which is very powerful and much more convenient than direct numerics.  To compare the two schemes, I consider a two-component mixture in two dimensions.  Fig. (\ref{fig:k}) shows the $ZT$ value of the composite as I vary the interface electrical conductance $\bar{\sigma}_{\rm int}$ and phonon thermal conductance $K_{\rm int}$.  In this case, I let $S_1=S_2=S_{\rm max}$, so that $ZT$ enhancement is the result of increased phonon scattering.  Both schemes give similar results, although the edge scheme shows slightly greater $ZT$ enhancement.  In the region of $ZT$ enhancement, the interfacial conductance is not appreciably larger than the bulk, so that the temperature and voltage drops aren't strongly localized at the interface.  This enables the center scheme to represent the fields reasonably well.  Moreover the enhancement is due to blocking phonons, or a small $\kappa_{\rm eff}^\gamma$.  Adding the large bulk $\kappa^\gamma_{\rm bulk}$ with the small $\kappa_{\rm int}^\gamma$ in series {\it a priori} results in an effective $\kappa^\gamma$ that's still small.  (For conductors in series, the smallest conductance dominates).  I therefore conclude that the approach adopted in the paper works well to describe $ZT$ enhancement via phonon scattering at the interface.

\begin{figure}[h!]
\begin{center}
\vskip 0.2 cm
\includegraphics[width=3.4in]{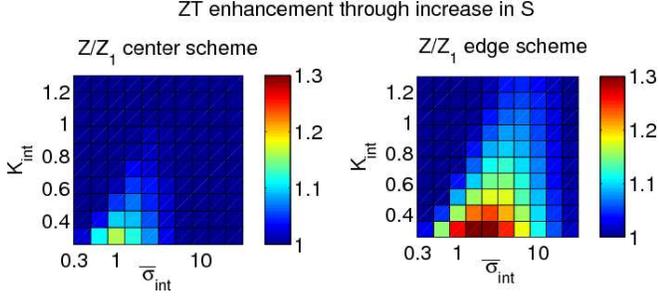}
\vskip 0.2 cm \caption{$ZT$ of the composite versus interface $\bar{\sigma}_{\rm int}$ and $K_{\rm int}$.  The system parameters are: $40{\rm x}40$ plaquettes in 2-d, $N_1=N_2=0.5$, $K_1=K_2=0.5$ (so that $Z_1 T=Z_2 T = 0.5$), $S_{\rm int} = 0.9~S_{\rm max}$.  These parameters lead to the same $ZT_{\rm int}$ as in Fig. (\ref{fig:k}).  Further analysis of the data shows that the center scheme underestimates the increase in $N$ (equivalently $S$) of the composite, resulting in a smaller $ZT$ relative to the edge scheme.}\label{fig:n}
\end{center}
\end{figure}

Fig. (\ref{fig:n}) shows $ZT$ as a function of interface properties for the two schemes when the bulk thermopower is small ($S_1=S_2=0.5 S_{\rm max}$).  The role of the interface in $ZT$ enhancement is to provide energy filtering of the electrons, increasing $S$ of the composite.  The two schemes' results are now rather different - the center scheme underestimates the $ZT$ enhancement by a notable margin.  This is because energy filtering is accomplished with a sharp temperature drop across the interface, which is not represented in the center scheme.  Moreover, adding the low bulk value of $\left(S\sigma\right)_{\rm bulk}$ in series with the high interface $\left(S\sigma\right)_{\rm int}$ {\it a priori} leads to a small effective $\left(S\sigma\right)$ (again, when adding these ``conductances" in series, the smallest one dominates); the potential increase in $S\sigma$ is partially nullified by the model construction.

\begin{thebibliography}{00}
\bibitem{footnote}
The equation of continuity for heat current is generally $\nabla \cdot j_Q = j \cdot \nabla V + j_Q \cdot \nabla T/T$.  In linear response the right-hand-side of this equation can generally be ignored. The full nonlinear version with heating was solved in the numerical calculation, and was found to agree with the results when
heating is neglected.
\bibitem{footnote1}
One reason for the insensitivity of $K_{\rm int}^{\rm max}$ to {\it how} $Z_2T$ is decreased is the fact that the absolute differences in parameters along different paths of $Z_2T$ degradation are relatively small.  For example, for $Z_2T = 0.375$, the ``bad $N_2$" case has $\left(\sigma_2,\kappa_2,S_2\right)=(1,3,0.9)$, while the ``bad $K_2$" has $\left(\sigma_2,\kappa_2,S_2\right)=(1,3.6,1)$.  Generally $K_{\rm int}^{\rm max}$ is slightly greater for the ``bad $K_2$" case.  This makes sense, as here the deficiency of material 2 (high conduction of phonons) is more directly addressed by the interface scattering.
\end{thebibliography}
\end{document}